\documentclass[10pt]{article}
\usepackage{graphicx}
\begin{document}
\title{Path integral evaluation of the one-loop effective potential in
field theory of diffusion-limited reactions}
\author{David Hochberg\footnote{Centro de Astrobiolog\'{\i}a (CSIC-INTA), Ctra.
Ajalvir Km. 4, 28850 Torrej\'{o}n de Ardoz, Madrid, Spain; email: hochberg@laeff.esa.es.
Corresponding author.} and
M.-P. Zorzano\footnote{Centro de Astrobiolog\'{\i}a (CSIC-INTA), Ctra.
Ajalvir Km. 4, 28850 Torrej\'{o}n de Ardoz, Madrid, Spain; email: zorzanomm@inta.es}}

\maketitle
\begin{abstract}
The well-established effective action and effective potential
framework from the quantum field theory domain is adapted and successfully
applied to classical field theories of the Doi and Peliti type
for diffusion controlled reactions. Through a number of benchmark
examples, we show that the direct calculation of the effective
potential in fixed space dimension $d=2$ to one-loop order
reduces to a small set of simple elementary functions, irrespective
of the microscopic details of the specific model. Thus the technique, which
allows one to obtain with little additional effort, the potentials
for a wide variety of different models, represents an important
alternative to the standard model dependent diagram-based  calculations.
The renormalized effective potential, effective equations of motion and
the associated renormalization group equations are computed in $d=2$ spatial dimensions for
a number of single species field theories of increasing complexity.\\ \\
\textbf{KEY WORDS}: Effective potential, renormalization group, reaction, diffusion.
\end{abstract}

\newpage

\section{\label{sec:intro}Introduction}

The effective potential provides physical insight into the behavior of
non-equilibrium stochastic dynamics, and takes into account both
nonlinear interactions and random fluctuations to a given number
of loops \cite{HPMVb}.  It has been used to study transitions in
non-equilibrium steady states \cite{Oerding} by means of the
renormalization group improved equation of state \cite{HK}, and
has been used most recently in nonperturbative renormalization group (RG)
methods applied to nonequilibrium critical phenomena \cite{Canet}.

Given the importance of the effective potential in statistical
physics, it is useful to have efficient analytic calculational
methods for its evaluation that are applicable to a wide variety
of non-equilibrium field theories.
In this paper, we take the well-established effective action and
effective potential formalism from quantum field theory and adapt it
to obtain renormalized closed form analytic expressions for the perturbative
potential and attendant renormalization group equations (RGE).
Once this is carried in detail for one
given model, it is easy to extend it to a wide class of different reaction
models with little additional effort.

To illustrate and validate these
techniques, we will consider some well studied field theories of
reaction and diffusion processes that have been derived from the
microscopic master equations following the procedures of Doi
\cite{Doi} and Peliti \cite{Peliti}. We will show by explicit
example that the method allows one to obtain both renormalized
potentials and RGEs with little effort. In fixed dimension, this represents
a major advantage over the delicate diagram based approach
where each separate reaction model requires defining and handling a distinct set
of diagrams and the working out of combinatorics.

We now outline the basic steps that take us from a given bare
action to the one-loop effective potential. The field theories we are interested
in are characterized by bare (unrenormalized) actions $S[\phi,{\bar
\phi}]$ that depend on a set of fields $\phi_i$, an equal number
of response fields ${\bar \phi}_i$, where $i = (1,2,...,N)$ and
$N$ is the number of independent species in the model. From $S$ we construct
the following array of second order functional derivatives of the
action:
\begin{equation}\label{smatrix}
S^{(2)}(\phi,{\bar \phi};x,y) =  \left(%
\begin{array}{cc}
 \frac{\delta^2}{\delta \bar \phi_i(x,t) \delta \bar \phi_j(y,s)}  &
\frac{\delta^2}{\delta \bar \phi_i(x,t) \delta \phi_j(y,s)}\\
 \frac{\delta^2}{\delta \phi_i(x,t) \delta \bar \phi_j(y,s)}  &
 \frac{\delta^2}{\delta \phi_i(x,t) \delta \phi_j(y,s)}\\
\end{array}
\right)S[\phi,{\bar \phi}],
\end{equation}
where $i,j = (1,2,\ldots,N)$. This is written in $2\times 2$ block
form. Since the potential is defined for constant and steady field
configurations, we easily evaluate the Fourier transform (FT) of
Eq. (\ref{smatrix}) via
\begin{equation}\label{Fourier}
FT\Big(S^{(2)}(\phi,{\bar \phi};x,y) \Big|_{\phi,{\bar \phi} =
const.}\Big) = M(k,\omega; \phi,{\bar \phi})\times
(2\pi)^{d+1}\delta^d(k-p)\delta(\omega - \Omega),
\end{equation}
where $M$ is a $2N \times 2N$ matrix and $d$ is the spatial
dimension. We then evaluate the determinant of $M$ and insert it
into the expression for the one-loop potential (see the Appendix
for a self-contained derivation):
\begin{equation}\label{effpot}
\mathcal{V}(\phi,{\bar \phi}) = V_0(\phi,{\bar \phi}) +
\frac{\mathcal{A}}{2} \int \frac{d^dk}{(2\pi)^d}\,
\int_{-\infty}^{\infty}\frac{d\omega}{2\pi} \ln\Big( \frac{\det
M(k,\omega; \phi,{\bar \phi})}{\det M(k,\omega;0,1)} \Big).
\end{equation}
Thus, given an action $S[\phi,{\bar \phi}]$, it is straightforward
to calculate the corresponding matrix $M$, Eq. (\ref{Fourier}),
followed by its determinant, and thence to the effective potential
Eq. (\ref{effpot}). The zero loop contribution $V_0$ corresponds
to the non-derivative part of the action $S$ (see Eq.
(\ref{definition})). When this is renormalized, we will refer to
it as the tree-level potential. The one loop contribution is
proportional to $\mathcal{A}$. The indicated integrations over
frequency $\omega$ and wavevector $k$ can be often be expressed entirely in
terms of elementary functions for fixed dimension. These points
will be demonstrated explicitly in what follows.

Once the effective potential is calculated to some order in
perturbation theory or loops, it is a simple matter to obtain the associated
RGEs \cite{HPMVa}. This procedure for obtaining the RGE's is based
on the so-called derivative technique, which was first introduced
in the work of Fujimoto et al. \cite{FRP} and subsequently
extensively developed in \cite{GLPQ}. The important observation is
that the bare theory does not depend on the arbitrary scale $\mu$
introduced by the renormalization scheme. This observation leads
to a simple identity. This identity yields an equation which is a
polynomial in the background field whose \textit{coefficients} are
precisely the RGE's that we seek. This is a very simple and powerful technique
that deserves wider use. Equally noteworthy is the fact that to
one-loop order, the case we treat here, the analytic calculation
of the effective potential is entirely straightforward and one can
dispense with model-dependent diagrammatic manipulations. This has
the clear advantage of allowing one to survey and analyze
comparatively a wide variety of reaction-diffusion models by
making simple algebraic substitutions to some of the terms
appearing within the \textit{universal} formula for the one-loop
effective potential. Knowledge of the effective potential also
allows one to reconstruct the effective action (up to wavefunction
renormalization) and deduce the corresponding equations of motion
obeyed by the physical density field. These equations of motion
are deterministic, yet incorporate the presence of the internal
reaction noise through modified effective force terms.

The remainder of this paper is concerned with the explicit
evaluation of Eq. (\ref{effpot}) for a selected number of single
species ($N=1$) diffusion limited reactions in $d=2$, and is
organized as follows. In the next Section, we calculate the
one-loop effective potential corresponding to pure pair
annihilation $A + A \rightarrow \emptyset$. This is renormalized
and the resultant finite expression is examined with and without
the loop corrections to reveal qualitatively the role of
fluctuations on the asymptotic states of the system. With the
potential calculated, it is a simple matter to obtain the
associated renormalization group equation (RGE) that controls the
scale dependence of the annihilation reaction rate. We also derive
the effective equation of motion satisfied by the fluctuation
averaged particle density. In Sec. \ref{sec:Gribov}, we calculate
the effective potential for the Gribov process which is described
by the set of single species reactions $A \rightarrow \emptyset, A
\rightarrow A + A, A + A \rightarrow A$ and extract the one-loop
RGE's and derive the effective equation of motion. Reaction and
diffusion systems with the underlying reaction process $A + A
\rightarrow \emptyset$ and $A \rightarrow (m+1)A$ are known as
branching and annihilating random walks (BARW). Here we calculate
the one-loop effective potential for arbitrary even $m$ in Sec.
\ref{sec:BARW} and extract the RGE's in Sec. \ref{sec:BARWRGE}. We
then specialize to the case $m=2$, solve the corresponding RGE's
and obtain the equation of motion for the noise-averaged density
field. The purpose of this paper is to carry these steps out explicitly in
detail in order to illustrate the techniques involved and the
ease with which they are applied to a variety of reaction models.
Conclusions and discussion are presented in Sec. \ref{sec:disc}.

\section{\label{sec:pair}Pair annihilation reaction $A+A \rightarrow \emptyset$}

Perhaps the simplest of all two-body irreversible recombination
processes is the single species pair annihilation reaction $A + A
\rightarrow {\rm inert}$. In this model, density fluctuations were
shown to alter radically the asymptotic decay laws predicted from
the mean-field rate equation approach for all dimensions $d\leq 2$
through a series of dimensional considerations, heuristic and
scaling arguments and computer simulations \cite{TW, KR}. The
anomalous scaling in the density was firmly established by Peliti
on the basis of field theoretical descriptions using the
renormalization group (RG)\cite{Peliti86}. Subsequent works
applying the RG to this system include \cite{Ohtsuki,DrozSas,FLS}.
An RG calculation for the reaction $kA \rightarrow {\rm inert}$ is
carried out in \cite{BPLee}, which recovers Peliti's result for
$k=2$. To illustrate the techniques mentioned above, we present a
detailed calculation of the one-loop approximation to the
effective potential in this simplest possible case. We renormalize
and examine this potential, use it for a simple derivation of
the corresponding RGE's in $d=2$, and to obtain the effective
equation of motion for the noise-averaged particle density.

\subsection{\label{sec:pairpot} One loop effective potential}

The following bare action will be the starting point for our effective potential
calculation:
\begin{equation}\label{spair}
S[\psi,\bar \psi] = \int d^dx \, dt \left[ \bar\psi
\Big(\partial_t +\xi - D\nabla^2 \Big)\psi - \lambda\big(1-
{\bar\psi}^2 \big)\psi^2 \right].
\end{equation}
The continuum parameters $D,\lambda$ appearing in the action Eq.
(\ref{spair}) stand for particle diffusion and the annihilation
reaction rate, respectively. A field-theoretic action $S$ corresponding to the pair
annihilation reaction including diffusion was first written down
almost twenty years ago by Peliti \cite{Peliti86}, but without the inclusion of the \emph{bare}
parameter $\xi$. This  deserves some special mention.
Physically, this is a mass term, and is normally taken identically to zero from the
outset in field-theoretic treatments of pair-annihilation. Here,
we must explicitly add it to the bare tree-level action Eq.
(\ref{spair}), since, as we will demonstrate below, it is required
in order to carry out consistently the one-loop renormalization of
the effective potential. Nevertheless, after this renormalization
is performed, we are free to set the finite and renormalized value
of $\xi$ to zero, and we will do so. Our renormalized action agrees with the
renormalized action obtained by other means \cite{Peliti86}.

We next carry out the basic steps outlined above in Sec
(\ref{sec:intro}) that take us from the bare action to the formal
expression of the one-loop potential. The matrix Eq.
(\ref{smatrix}) of second functional derivatives of the action is
\begin{eqnarray}\label{spairJacobi}
S^{(2)}(\phi,{\bar \phi}) &=&
\left(%
\begin{array}{cc}
  2\lambda \phi^2 & -\partial_t + \xi - D\nabla^2 + 4\lambda \bar \phi \phi \\
  \partial_t +\xi - D\nabla^2 + 4\lambda \bar \phi \phi & -2\lambda(1 - {\bar \phi}^2) \\
\end{array}
\right)\nonumber \\
&\times&\delta^d(x-y)\delta(t-s).
\end{eqnarray}
This is a tensor product of a $2 \times 2$ matrix in field space
times an infinite dimensional but diagonal array in configuration
space and in time (the product of delta functions). Since the
effective action and potential depend on the \textit{averages} of
the fluctuating fields, we must distinguish between the pairs of
fields $\phi$, $\psi$, and $\bar \phi$,$\bar \psi$, respectively;
see Eq. (\ref{averages}). For constant fields we pass to Fourier
variables to obtain the array $M$, Eq. (\ref{Fourier}), where
\begin{equation}\label{matrix}
M(k,\omega; \phi, \bar \phi) = \left(%
\begin{array}{cc}
  2\lambda \phi^2 & -i\omega + \xi + Dk^2 + 4\lambda \bar \phi \phi \\
  +i\omega + \xi + Dk^2 + 4\lambda \bar \phi \phi & -2\lambda(1 - {\bar \phi}^2) \\
\end{array}
\right).
\end{equation}
Inserting this matrix into the expression for the one-loop
effective potential Eq. (\ref{effpot}) yields (recall that $V_0 =
\frac{S}{VT}$ for constant fields)
\begin{eqnarray}\label{pairpot}
& &\mathcal{V}(\phi,\bar \phi) = -\lambda\phi^2(1 - {\bar \phi}^2) +
\xi \bar\phi \phi \nonumber \\
&+& \frac{\mathcal{A}}{2}\int
\frac{d^dk}{(2\pi)^d}\,
\int_{-\infty}^{\infty}\frac{d\omega}{2\pi} \ln\Big(
\frac{\omega^2 + [Dk^2 + \xi + 4\lambda\phi \bar \phi]^2 +
4\lambda^2\phi^2(1 - {\bar \phi}^2)}{\omega^2 + [Dk^2 +
\xi]^2}\Big).\nonumber \\
&&
\end{eqnarray}
We first carry out the frequency integral, which can be evaluated
exactly in closed form. By means of the following identity (see
Eq.(4.222.1) of \cite{GR}) valid for $C,A > 0$
\begin{equation}\label{freqint}
\int_{-\infty}^{\infty} d\omega \ln\Big(\frac{\omega^2 +
C^2}{\omega^2 + A^2}\Big) = 2\pi \big(C - A),
\end{equation}
we can write the one-loop $O(\mathcal{A})$ contribution in Eq.
(\ref{pairpot}) as follows:
\begin{eqnarray}\label{Iunreg}
\mathcal{I} &=& \frac{1}{2} \int_0^{\infty} k^{d-1} dk\, \int
\frac{d\Omega_d}{(2\pi)^d}\, \left\{ \sqrt{(Dk^2 + \xi + 4\lambda
\phi \bar \phi)^2 + 4\lambda^2\phi^2(1 - {\bar \phi}^2)} \right. \nonumber \\
&-& \left. \sqrt{(Dk^2+ \xi)^2} \right\},
\end{eqnarray}
where $d\Omega_d$ is the element of solid angle in $d$-dimensional
wavevector space. We now temporarily regulate the integral over
wavenumber modulus $k$ with an ultraviolet cut-off $k  < \Lambda$
and evaluate it directly in $d=2$ dimensions. Introducing the
change of variable $u = Dk^2$, we can express the regulated $d=2$
integral Eq. (\ref{Iunreg}) as follows:
\begin{equation}\label{Icutoff}
\mathcal{I}_{\Lambda} = \frac{1}{8\pi D} \int_{u = 0}^{u =
D\Lambda^2} du \left\{ \sqrt{u^2 + a_1u + a_2} - \sqrt{(u +
\xi)^2} \right\},
\end{equation}
where
\begin{eqnarray}\label{a1}
a_1 &=& 2(\xi + 4\lambda \phi \bar\phi) \\\label{a2} a_2 &=&
\frac{{a_1}^2}{4} +  4\lambda^2\phi^2(1 -
{\bar\phi}^2)\\\label{delta} \Delta &=& 4a_2 - a_1^2 = 16\lambda^2
\phi^2(1 - {\bar\phi}^2),
\end{eqnarray}
and which can be expressed in terms of the following elementary
functions (see e.g., (2.262.1) and (2.261) in \cite{GR}):
\begin{eqnarray}\label{Iregulated}
(8\pi D)\mathcal{I}_{\Lambda} &=&  \frac{(2D\Lambda^2 +
a_1)}{4}\sqrt{a_2 + a_1D\Lambda^2 + (D\Lambda^2)^2} \\ \nonumber
&+& \frac{\Delta}{8}\ln\Big(2\sqrt{a_2 + a_1D\Lambda^2 +
(D\Lambda^2)^2} + 2D\Lambda^2 + a_1\Big)\\ \nonumber &-&
\frac{1}{2}(D\Lambda^2 + \xi)^2 -\frac{a_1}{4}\sqrt{a_2} +
\frac{\xi^2}{2} -\frac{\Delta}{8}\ln\big(2\sqrt{a_2} + a_1\big).
\end{eqnarray}
A careful asymptotic analysis of the regulated loop integral Eq.
(\ref{Iregulated}) reveals that $\mathcal{I}_{\Lambda}$ contains a
quadratic and a logarithmic divergence in $\Lambda$ in addition to
certain finite terms as this UV cutoff is taken to infinity,
namely
\begin{eqnarray}\label{Iregulated2}
\lim_{\Lambda \rightarrow \infty} \mathcal{I}_{\Lambda} &=&
\frac{1}{8\pi D}\left\{ \big(\frac{a_1}{2} - \xi \big)
(D\Lambda^2) - \frac{\Delta}{8}\ln\Big(\frac{2\sqrt{a_2} +
a_1}{4D\Lambda^2} \Big) + \frac{1}{16}(\Delta + 2a_1^2) \right. \nonumber \\
&-& \left. \frac{a_1}{4}\sqrt{a_2} \right\} + O(\Lambda^{-2}).
\end{eqnarray}
A particularly useful separation between finite and divergent
pieces is obtained by introducing an \textit{arbitrary} finite
momentum scale $\mu$ and writing the argument of the logarithm as
follows:
\begin{equation}\label{useful}
\frac{2\sqrt{a_2} + a_1}{4D\Lambda^2} = \frac{2\sqrt{a_2} +
a_1}{4D\mu^2}\times\frac{\mu^2}{\Lambda^2}.
\end{equation}
After using this simple identity Eq. (\ref{useful}), we have the
following regulated one loop effective potential Eq.
(\ref{pairpot}) in $d=2$ space dimensions:
\begin{eqnarray}\label{Vregulated}
\mathcal{V}(\phi,\bar \phi) &=& -\lambda\phi^2(1-{\bar \phi}^2)
+ \xi \phi \bar \phi \nonumber \\
&+& \frac{\mathcal{A}}{8\pi D}\left\{ \big(\frac{a_1}{2} - \xi
\big) (D\Lambda^2) + \frac{\Delta}{8}\ln \frac{\Lambda^2}{\mu^2} -
\frac{\Delta}{8}\ln\Big(\frac{2\sqrt{a_2} + a_1}{4D\mu^2} \Big) \right. \nonumber \\
&+& \left. \frac{1}{16}(\Delta + 2a_1^2) -\frac{a_1}{4}\sqrt{a_2} \right\} +
O(\Lambda^{-2}).
\end{eqnarray}
This expression is given in terms of bare (unrenormalized)
parameters $(\lambda,\xi)$ and the dependence on the UV cutoff
$\Lambda$ is explicit. As pointed out above, the tree-level term
$\xi \phi \bar \phi$ will be needed for the renormalization
program. Note that the regulated effective potential Eq.
(\ref{Vregulated}), is seen by inspection (due to elementary
properties of the logarithm function) to be manifestly independent
of the arbitrary scale $\mu$. This is the key observation needed
in order to derive the RGE's, of which we will make direct use of
below.

We now proceed to renormalize, i.e., absorb the divergences into
the bare parameters $\lambda$ and $\xi$. To carry this out, we
write
\begin{eqnarray}\label{AAcounter1}
\lambda &=& \lambda(\mu) + \mathcal{A}\kappa_1 + O(\mathcal{A}^2)\\
\label{AAcounter2}
\xi &=& \xi(\mu) + \mathcal{A}\kappa_2 +
O(\mathcal{A}^2),
\end{eqnarray}
in which we decompose each bare parameter into a finite
renormalized and possibly $\mu$-dependent part plus a divergent
counterterm. Note the divergent counterterms start off at order
$O(\mathcal{A})$, which is the same order at which the divergences
in the regulated potential appear. These counterterms
$\kappa_1,\kappa_2$ are chosen to render the one-loop effective
potential finite and UV-cutoff independent. Note the very
important fact that the stochastic fluctuations induce divergent
terms at the one-loop level whose coefficients have
\emph{identical} polynomial field structure (namely,
$\frac{a_1}{2} - \xi$ and $\Delta$, see Eq. (\ref{a1}) and Eq.
(\ref{delta})) as those appearing in the bare tree-level potential
$V_0$. Inserting Eq. (\ref{AAcounter1},\ref{AAcounter2}) into Eq.
(\ref{Vregulated}) yields the solution for the two counterterms:
\begin{eqnarray}\label{AAcounterterms1}
\kappa_1 &=& +\frac{\lambda^2(\mu)}{4\pi D} \ln \frac{\Lambda^2}{\mu^2} \\
\label{AAcounterterms2}
\kappa_2 &=& -\frac{\lambda(\mu)}{2\pi}
\Lambda^2.
\end{eqnarray}
In this final step, and after cancelling off the UV divergences,
we obtain the finite and renormalized one-loop effective potential
for pair-annihilation (in $d=2$):
\begin{eqnarray}\label{Vrenorm}
\mathcal{V}(\phi,\bar \phi) &=& -\lambda(\mu)\phi^2(1-{\bar \phi}^2)
\nonumber \\
&+& \frac{\mathcal{A}}{8\pi D} \Big\{-
2\lambda^2(\mu)\phi^2(1-{\bar \phi}^2)\ln
\Big(\frac{\lambda(\mu)\phi\sqrt{1+3{\bar \phi}^2} +
2\lambda(\mu)\phi \bar \phi}{D \mu^2} \Big)  \nonumber \\
&+&  \lambda^2(\mu) \phi^2(1- {\bar \phi}^2) +
8\lambda^2(\mu)\phi^2{\bar \phi}^2 - 4\lambda^2(\mu)\phi^2 \bar
\phi\sqrt{1 + 3{\bar \phi}^2}\,\, \Big\} \nonumber \\
&+& O(\mathcal{A}^2).
\end{eqnarray}
We are now free to assign zero to the finite \emph{renormalized}
value of $\xi$, and have done so here and in the following
subsections. Below, when we come to discuss the RGE's, we will see
that the renormalized parameter $\xi$ does \emph{not} run with
scale, at least to one-loop order, proving that this is a valid
choice. In this expression Eq. (\ref{Vrenorm}) it is understood
that $\lambda \equiv \lambda(\mu)$ is the renormalized,
$\mu$-dependent coupling and we have set the finite renormalized
value of $\xi(\mu)$ identically to zero. We next determine their
scale dependences.

\subsection{\label{sec:pairRGE}The one-loop RGEs}

Since the bare theory Eq. (\ref{Vregulated}) does not depend on
the arbitrary scale $\mu$ introduced by the renormalization scheme
(recall our use of the trivial identity in Eq. (\ref{useful})), we
are able to derive the renormalization group equation that the
coupling $\lambda(\mu)$ must satisfy, from the identity
\cite{HPMVa,FRP,GLPQ}
\begin{equation}\label{RGEa}
\mu \frac{d \mathcal{V}(\phi,\bar \phi)}{d \mu } = 0,
\end{equation}
which upon inserting Eq. (\ref{Vrenorm}) immediately yields the
RGE. Differentiating as indicated, we get
\begin{eqnarray}\label{RGEb}
0 &=& \mu\frac{d}{d\mu}\Big(-\lambda\phi^2(1-{\bar \phi}^2)\Big)\nonumber \\
&+&
\frac{\mathcal{A}}{8\pi D}\mu\frac{d}{d\mu}\Big(-
2\lambda^2\phi^2(1-{\bar \phi}^2)\ln
\big(\frac{\lambda\phi\sqrt{1+3{\bar \phi}^2} +
2\lambda\phi \bar \phi}{D \mu^2} \big) \nonumber \\
&+& \lambda^2 \phi^2(1- {\bar \phi}^2) + 8\lambda^2\phi^2{\bar
\phi}^2 - 4\lambda^2\phi^2 \bar \phi\sqrt{1 + 3{\bar \phi}^2}\Big)
+ O(\mathcal{A}^2),
\end{eqnarray}
which implies
\begin{equation}\label{RGEc}
\mu\frac{d}{d\mu}\Big(\lambda\phi^2(1-{\bar \phi}^2)\Big) =
\frac{\mathcal{A}}{2\pi D}\lambda^2\phi^2(1-{\bar \phi}^2) +
O(\mathcal{A}^2).
\end{equation}
Note that derivatives with respect to $\mu$ of the other terms on
the right hand side of Eq. (\ref{RGEb}) proportional to
$\mathcal{A}$ lead to contributions at the next higher order
$O(\mathcal{A}^2)$, and so can be neglected at the one-loop order
we are working at. Indeed, from Eq. (\ref{AAcounter1}) we see that
$d\lambda(\mu)/{d\mu} = d(\lambda - \mathcal{A}\kappa_1(\mu))/d\mu
= -\mathcal{A} d \kappa_1(\mu)/d\mu = O(\mathcal{A}^2)$.

Cancelling the \emph{common} field-dependent factor
$\phi^2(1-{\bar \phi}^2)$ from both sides of Eq. (\ref{RGEc})
yields the RGE for $\lambda$ in $d=2$:
\begin{equation}\label{RGEd}
\mu \frac{d \lambda}{d \mu } = \frac{\mathcal{A}}{2\pi
D}\lambda^2,
\end{equation}
valid to one-loop order. This can be solved immediately and
gives
\begin{eqnarray}\label{solnRGEd}
\lambda(\mu) &=& \lambda_0 \Big(1 -
\frac{\mathcal{A}\lambda_0}{2\pi D} \ln\frac{\mu}{\mu_0} +
O(\mathcal{A}^2) \Big)^{-1}\nonumber \\
& \approx& \lambda_0 \Big(1 + \frac{\mathcal{A}\lambda_0}{2\pi D}
\ln\frac{\mu}{\mu_0} \Big) + O(\mathcal{A}^2),
\end{eqnarray}
where $\lambda_0 \equiv \lambda(\mu_0)$, is the value of the
coupling at some initial arbitrary reference scale $\mu_0$.

If we retain the dependence on the renormalized $\xi$ in Eq.
(\ref{Vrenorm}) and deduce the consequences following from Eq.
(\ref{RGEa}), we now obtain two RGE's, corresponding to the two
independent polynomial terms $\phi^2(1-{\bar \phi}^2)$ and $\bar
\phi \phi$, respectively. Once again, we obtain Eq. (\ref{RGEd})
as well as
\begin{eqnarray}
\mu \frac{d \xi}{d \mu } &=& 0 + O(\mathcal{A}^2), \\
\xi(\mu) &=& \xi(0) + O(\mathcal{A}^2),\nonumber
\end{eqnarray}
indicating that the parameter $\xi$ does \emph{not} run with
scale, at least to one-loop order. So, it is consistent to take
the renormalized (but not the bare) mass parameter to be strictly
constant, and setting $\xi(\mu) = \xi(0) = 0$ is a perfectly valid
choice.

This RGE Eq. (\ref{RGEd}) has been calculated directly in $d=2$,
which coincides of course with the upper critical dimension of the
pair-annihilation process \cite{KR}. When working with the RG, it
is useful to have knowledge of the scale dependence in variable
dimension $d$. This would require us to go back and attempt to
calculate the potential also in arbitrary $d$, which is a
difficult task. This is because the one loop potential represents
an infinite sum of Feynman diagrams. While individual diagrams are
usually easy to analytically continue to arbitrary dimensions,
their infinite sums are not. However, we can gain some useful
insight into the structure that the variable-$d$ RGE must have by
using some simple dimensional analysis in conjuntion with the
$d=2$ RGE in Eq. (\ref{RGEd}). In this way, we arrive at the
approximate mathematical form of the RGE for the dimensionless
coupling, which may be expected to hold for values of  $d$ close
to $d=2$. We comment below in what sense it is approximate. In
regards the scaling dimensions of the field theory, the diffusion
constant $D$ can be absorbed into a rescaling of the time $t
\rightarrow Dt $. Doing so in the action Eq. (\ref{spair}) and
expressing the dimensions of the resultant terms in the action in
powers of momentum yields \cite{BPLee}
\begin{equation}\label{dims}
[t] = p^{-2} \qquad [\bar \psi] = p^0 \qquad [\psi] = p^d \qquad
[\xi/D] = p^2 \qquad [\lambda/D] = p^{2-d}.
\end{equation}

From the last equality in Eq. (\ref{dims}), we are lead to define
a dimensionless coupling $g$ in any $d$ via
\begin{equation}\label{dimlesscoupling}
g = \frac{\lambda}{D \mu^{\epsilon}},
\end{equation}
where $\epsilon = 2-d$.
We can write Eq. (\ref{RGEd}) so that it is \emph{dimensionally}
consistent for all dimensions $d$. This is achieved by first
noting that (we set $\mathcal{A} = 1$ in what follows)
\begin{equation}\label{RGEcca}
\mu \frac{d \lambda}{d \mu } = \Big(\frac{K_d }{D
\mu^{\epsilon}}\Big)_{d=2}\lambda^2,
\end{equation}
where
\begin{equation}\label{solidangle}
K_d = \int \frac{d\Omega_d}{(2\pi)^d}=
[\Gamma(d/2)\pi^{d/2}2^{d-1}]^{-1}
\end{equation}
is a geometric factor that results from the integration over solid
angles and which is part of the complete $d$-dimensional loop
integral in Eq. (\ref{Iunreg}). Note that $K_2 = \frac{1}{2\pi}$
and that $g(d=2) = \frac{\lambda}{D}$. We next extend Eq.
(\ref{RGEcca}) to all dimensions by writing
\begin{equation}\label{RGEcc}
\mu \frac{d \lambda}{d \mu } = \frac{K_d }{D
\mu^{\epsilon}}\lambda^2.
\end{equation}
This equation is of course only true in $d=2$, and the
approximation comes in using dimensional analysis to extend it to
any dimension.  Lastly, substituting Eq. (\ref{dimlesscoupling})
into Eq. (\ref{RGEcc}), and expressing $\lambda$ in terms of $g$,
yields the (admittedly approximate) one loop beta function for
$g$:
\begin{equation}\label{beta}
\beta(g) \equiv \mu \frac{d g}{d \mu } = -\epsilon g + K_d\, g^2.
\end{equation}
This beta function Eq. (\ref{beta}) is quadratic in $g$ and for
$d<2$ has the nontrivial fixed point $\beta(g^*) = 0$ at $g^* =
\frac{ \epsilon}{K_d} = O(\epsilon)$. This fixed point is of order
$\epsilon$. For $d>2$, $g^* = 0$. Inspection of Eq. (\ref{Iunreg})
shows that the calculation of the one-loop RGE for $ \lambda$
valid for arbitrary $d$ would require the integration of the
corresponding integrand over the modulus of $k$ with weight
$k^{d-1}$, in addition to the integration over solid angle Eq.
(\ref{solidangle}). The above result takes the latter in to
account, but the former is evaluated at fixed $d=2$, so we can
expect the overall numerical coefficient of the quadratic $g^2$
term to be in error in this respect. Nevertheless, for values of
$d$ sufficiently close to $d=2$, this should provide qualitatively
correct results. By way of comparison, the exact one-loop beta
function calculated in \cite{BPLee} is given by
\begin{equation}\label{Leebeta}
\beta(g) = -\epsilon g + \frac{\Gamma(2-d/2)}{2\pi} g^2.
\end{equation}
For $d>2$, $g^* = 0$. For $d<2$ the nontrivial fixed point is
given by $g^*_L = \frac{2\pi \epsilon}{\Gamma(2-d/2)}$.  At $d =
2$, these two beta functions Eq. (\ref{beta}) and Eq.
(\ref{Leebeta}) are identical, and we have $\beta(g) =
\frac{1}{2\pi}g^2$. For $d<2$ we can compare the deviation of the
approximate to exact nontrivial fixed point via the ratio
$g^*/g^*_L = \frac{\Gamma(2-d/2)}{2\pi K_d}$.

\subsection{\label{sec:paireom}Effective equation of motion}

\begin{figure}[h]
\begin{center}
\includegraphics[width=0.8  \textwidth]{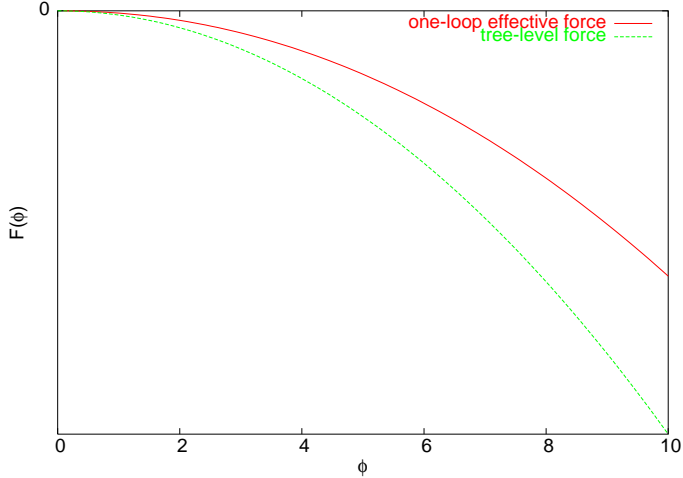}
\caption{\label{fig1} One-loop effective and
tree-level force $F(\phi)$. The zero at the origin corresponds to
the inactive state.}
\end{center}
\end{figure}

From the one-loop potential we can construct the effective action
to the same (one-loop) order via \cite{Jackiw,IIM,ZJ,Rivers,CW}
\begin{equation}\label{effactcons}
\Gamma[\phi,\bar \phi] = \int d^dx \, dt \, \left\{ Z^{1/2}{\bar
Z}^{1/2}\bar \phi(\partial_t - D\nabla^2)\phi + {\cal V}
(\phi(x,t),\bar \phi(x,t)) + \ldots \right\},
\end{equation}
where $Z$ and $\bar Z$ denote the wavefunction renormalization
constants for $\phi$ and $\bar \phi$, respectively, and the dots
represent higher derivative terms in the fields. From the fact
that there is no wavefunction renormalization for
pair-annihilation \cite{Peliti, BPLee}, we can immediately set $Z
= \bar Z = 1$ in Eq. (\ref{effactcons}). Then, from Eq.
(\ref{effeqom0}) and Eq. (\ref{Vrenorm}), we can derive the
(one-loop) effective equations of motion for pair-annihilation in
$d=2$ obeyed by the noise-averaged fields $\bar \phi(x,t) =
\langle \bar \psi(x,t) \rangle_{(0,0)}$ and $\phi(x,t) = \langle
\psi(x,t) \rangle_{(0,0)}$. We find that the equation of motion
for $\bar \phi$ (recall $N=1$)
\begin{equation}\label{eom1}
\Big(\frac{\delta \Gamma[\phi, \bar \phi]}{\delta
\phi}\Big)\Big|_{\phi, \bar \phi} = 0,
\end{equation}
is identically solved by the stationary and homogeneous solution
$\bar \phi = 1$. The equation of motion of the physical density
field $\phi$ follows from
\begin{equation}\label{eom2}
\Big(\frac{\delta \Gamma[\phi, \bar \phi]}{\delta \bar
\phi}\Big)\Big|_{\phi, \bar \phi = 1} = 0,
\end{equation}
which when written out explicitly, implies that
\begin{eqnarray}\label{paireom}
\big(\partial_t - D\nabla^2\big)\phi &=& -\frac{\partial
\mathcal{V}}{\partial \bar{\phi}}\Big|_{\bar \phi = 1}\\ \nonumber
&=& -2\lambda \phi^2 - \frac{\cal{A}}{2\pi D}\lambda^2\phi^2
\ln\Big(\frac{4\lambda \phi}{D\mu^2}\Big) + O({\cal A}^2)\\
\nonumber
&=& F_{\phi}.
\end{eqnarray}
Here, the effective force experienced by the field $\phi$ is
denoted by $F_{\phi}=-\frac{\partial \mathcal{V}}{\partial
\bar{\phi}}$. In Fig. \ref{fig1} we show a qualitative
example of the force experienced by the density field $\phi$.
Notice that since the recovery force is less strong in the
renormalized case (the fluctuations cause the slope at the origin
to be less negative, relative to the tree-level case), the rate of
approach to the origin will be slower. Putting $\mathcal{A} = 0$
on the right hand side of Eq. (\ref{paireom}) yields the so-called
rate equation which one might write down as a first approximation
to the equation of motion for the density. If we let $n(t)$ be the
spatial average of $\phi$, then we see that the only stationary
solution of Eq. (\ref{paireom}) is $n = 0$, corresponding to the
inactive state with no particles remaining, which of course, comes
as no surprise. Zeroes of the effective force correspond to
maxima/minima of the effective potential. Therefore, since the
logarithm of a small number is negative, it appears as though the
one-loop correction in Eq. (\ref{paireom}) has turned the minimum
at the origin into a maximum and caused a new minimum to appear
away from the origin, corresponding to a putative noise induced
active state. However, it is wise to proceed with caution in this
specific example: the apparent new minimum occurs at a value of
$n$ determined by (note from Eq. (\ref{dims}) that the ratios
$\lambda/D$ and $n/\mu^2$ are pure dimensionless numbers in $d=2$)
\begin{equation}\label{newsolnpair}
\Big(\frac{\lambda}{D}\Big) \ln\Big(\frac{4\lambda n}{D
\mu^2}\Big) = -4\pi.
\end{equation}
Since we expect higher orders in perturbation theory to bring in
higher powers of $(\lambda/D) \ln\big(4\lambda n/D \mu^2\big)$,
this new state lies well outside the expected range of validity of
the one-loop approximation \cite{CW}, even for arbitrarily small
reaction rates (or large diffusions) $\lambda/D$. This new
solution Eq. (\ref{newsolnpair}) must be rejected as an artefact
of the approximation \footnote{The one-loop effective \emph{force}
in the pair-annihilation model is mathematically similar to the
one-loop potential for the massless quartically self-interacting
meson field, for which an analogous criticism applies. This point
has been taken over and adapted to the context of the present
paper; see Ref. \cite{CW}. }. Higher loops are needed to assess
the situation. However, this same criticism can not be applied to
the effective force associated with the Gribov process, the next
example to be considered below.

\section{\label{sec:Gribov}Gribov process $A \rightarrow \emptyset,
A \rightarrow A + A, A + A \rightarrow A$}

As a second application of the methods outlined in Sec
(\ref{sec:intro}), we consider the competing reactions of single
particle annihilation, splitting and two-body recombination.  This
is a simple model exhibiting a non-equilibrium phase transition
between active and inactive or absorbing stationary states in
which the stochastic fluctuations cease entirely. A mathematical
equivalence between such chemical processes and Reggeon field
theory has been firmly established \cite{GS,J}. Below we calculate
the corresponding one-loop potential, carry out the
renormalization and obtain and solve the RGE's for the rates of
pair recombination and fission directly in $d=2$ dimensions.

\subsection{\label{sec:Gribovpot} One loop effective potential}

The following field-theoretic bare (unrenormalized) action $S$
will be the starting point for our effective potential calculation:
\begin{eqnarray}\label{sGribov}
S[\psi,\bar \psi] &=& \int d^dx \, dt \left[ \bar\psi
\Big(\partial_t + \xi - D\nabla^2 \Big)\psi + \sigma\big(1-
{\bar\psi}\big)\bar \psi \psi -r \big(1- {\bar\psi}\big)\psi \right. \nonumber \\
&-& \left. \lambda\big(1- {\bar\psi} \big){\bar \psi}\psi^2 + b\bar \psi +
a\right].
\end{eqnarray}
The continuum parameters $D,\lambda$ appearing in this action
stand for particle diffusion and the reaction rate for two-body
recombination ($A + A \rightarrow A$), respectively. The parameter
$\sigma$ is the rate of reproduction or fission ($A \rightarrow A
+ A$) and $r$ is the particle decay rate ($A \rightarrow
\emptyset$). Similarly to the pair-annihilation case worked out
above, the additional bare terms parametrized by $\xi$, $b$ and
$a$ that we have introduced in Eq. (\ref{sGribov}), will be
required in order to carry out consistently the one-loop
renormalization of the effective potential. Nevertheless, after
this renormalization is performed, we are free to set the finite
and \textit{renormalized} values of all these couplings
identically to zero: $\xi = b = a = 0$. We will so do in the final
renormalized expression for the effective potential obtained
below. Once these taken to be zero, the renormalized action (\ref{sGribov})
will correspond to the Gribov process \cite{Howard}.

Evaluating the array in ({\ref{smatrix}) for this action, taking
constant fields and then evaluating the Fourier transform yields
the following matrix (note this is independent of both $a$ and
$b$)
\begin{eqnarray}\label{sGribovJacobi}
M(k,\omega; \phi, \bar \phi)_{11} &=& -2\sigma\phi + 2\lambda \phi^2 \\
M(k,\omega; \phi, \bar \phi)_{12} &=&
-i\omega + \xi + Dk^2 + \sigma + r -2\sigma\bar \phi
-2\lambda\phi(1 - 2 \bar \phi) \\
M(k,\omega; \phi, \bar \phi)_{21} &=& M^*(k,\omega; \phi, \bar \phi)_{12} \\
M(k,\omega; \phi, \bar \phi)_{22} &=& -2\lambda(1 - {\bar \phi})\bar \phi.
\end{eqnarray}
Inserting this matrix into the expression for the one-loop
effective potential Eq. (\ref{effpot}) yields
\begin{eqnarray}\label{Gribovpot}
\mathcal{V}(\phi,\bar \phi) &=& \sigma(1-\bar \phi)\bar \phi \phi
-r\phi(1 -\bar \phi) -\lambda(1 - \bar \phi)\bar \phi \phi^2 +
\xi\bar \phi \phi
+ b\bar \phi + a \nonumber \\
&+& \frac{\mathcal{A}}{2}\int \frac{d^dk}{(2\pi)^d}\,
\int_{-\infty}^{\infty}\frac{d\omega}{2\pi}
\ln\Big( \frac{\det M(k,\omega; \phi,{\bar \phi})}{\det M(k,\omega;0,1)} \Big),
\end{eqnarray}
where,
\begin{eqnarray}
-\det M(k,\omega; \phi,{\bar \phi})&=&
\omega^2 + [Dk^2 + \xi + \sigma + r -2\sigma\bar \phi
-2\lambda\phi(1 - 2\bar \phi)]^2 \nonumber \\
&+& 2\lambda\bar
\phi(1-\bar\phi)(-2\sigma\phi + 2\lambda\phi^2),\\
-\det M(k,\omega;0,1) &=&
{\omega^2 + [Dk^2
+ \xi - \sigma + r]^2}.
\end{eqnarray}
The frequency integration in Eq. (\ref{Gribovpot}) is again
performed immediately using Eq. (\ref{freqint}). We regulate and
evaluate the remaining wavenumber integral directly in $d=2$ as
above in Sec. (\ref{sec:pairpot}). The integral that is needed
is similar in structure to that worked out in Eq.
(\ref{Icutoff}). After some algebra, we find that the UV-regulated
integral takes a form identical to that in Eq.
(\ref{Iregulated2}). Put $f = \xi + r + \sigma$, then for the bare
action Eq. (\ref{sGribov}) we have $I_{\Lambda}$ as written in Eq.
(\ref{Iregulated2}), where now the coefficients turn out to be
\begin{eqnarray}
a_1 &=& 2\big(f - 2\sigma \bar \phi -2\lambda \phi(1-2\bar
\phi) \big) \\
a_2 &=& [f - 2\sigma \bar \phi -2\lambda \phi(1-2\bar \phi)]^2 +
2\lambda\bar \phi(1-\bar \phi)(-2\sigma\phi +2\lambda
{\phi}^2) \\
\Delta &=& 4a_2 - a_1^2 = -16\lambda \sigma(1 - \bar \phi) \bar
\phi \phi + 16\lambda^2(1 - \bar \phi) \bar \phi \phi^2.
\end{eqnarray}
At this point we have the following regulated one loop effective
potential for the Gribov process (where $F = \xi-\sigma + r$)
\begin{eqnarray}\label{Vregulated2}
\mathcal{V}(\phi,\bar \phi) &=& \sigma(1-\bar \phi)\bar \phi \phi
-r\phi(1 -\bar \phi) -\lambda(1 - \bar \phi)\bar \phi \phi^2 + b
\bar \phi +\xi \phi\bar \phi + a \nonumber \\  &+&
\frac{\mathcal{A}}{8\pi D}\Big\{ \big(\frac{a_1}{2} - F \big)
(D\Lambda^2) + \frac{\Delta}{8}\ln \frac{\Lambda^2}{\mu^2} -
\frac{\Delta}{8}\ln\Big(\frac{2\sqrt{a_2} + a_1}{4D\mu^2} \Big)
\nonumber \\
&+&
\frac{1}{16}(\Delta + 2a_1^2) -\frac{a_1}{4}\sqrt{a_2} \Big\} +
O(\Lambda^{-2}).
\end{eqnarray}
This expression is given in terms of bare (unrenormalized)
parameters $(\sigma,r,\lambda,\xi,b,a)$ and the dependence on the
UV cutoff $\Lambda$ is explicit. As remarked above, the terms
proportional to $b\bar \phi$, $\xi \phi \bar \phi$ and the
constant $a$ will be needed for carrying out the renormalization
program. Note that the regulated effective potential Eq.
(\ref{Vregulated2}) is independent of the arbitrary scale $\mu$.

We now proceed to renormalize, i.e., absorb the divergences into
the bare parameters . Specifically, we write
\begin{eqnarray}
\label{Gcounter1}
\sigma &=& \sigma(\mu) + \mathcal{A}\kappa_1 +
O(\mathcal{A}^2) \\
\label{Gcounter2} r &=& r(\mu) + \mathcal{A}\kappa_2 +
O(\mathcal{A}^2)\\
\label{Gcounter3}
\lambda &=& \lambda(\mu) + \mathcal{A}\kappa_3 + O(\mathcal{A}^2)\\
\label{Gcounter4} \xi &=& \xi(\mu) + \mathcal{A}\kappa_4 +
O(\mathcal{A}^2)\\
\label{Gcounter5} b &=& b(\mu) + \mathcal{A}\kappa_5 +
O(\mathcal{A}^2)\\
\label{Gcounter6} a &=& a(\mu) + \mathcal{A}\kappa +
O(\mathcal{A}^2).
\end{eqnarray}
The divergent coefficients $\kappa_i$ of the counterterms are
chosen to cancel off the divergences in the one loop regulated
contribution Eq. (\ref{Vregulated2}). Inserting Eq.
(\ref{Gcounter1}-\ref{Gcounter6}) into Eq. (\ref{Vregulated2})
yields the solutions
\begin{eqnarray}\label{counterterms3}
\kappa_1 &=& \frac{\lambda(\mu) \sigma(\mu)}{4\pi D} \ln \frac{\Lambda^2}{\mu^2} \\
\kappa_2 &=& \kappa_4 = -\frac{\lambda(\mu)}{4\pi} \Lambda^2 \\
\kappa_3 &=& \frac{\lambda^2(\mu)}{4\pi D} \ln \frac{\Lambda^2}{\mu^2}  \\
\kappa_5 &=& -\kappa = \frac{\sigma(\mu)}{4\pi} \Lambda^2.
\end{eqnarray}
After this step, we obtain the finite and renormalized one-loop
effective potential for the Gribov process in $d=2$:
\begin{eqnarray}\label{VGribov}
\mathcal{V}(\phi,\bar \phi) &=& \sigma(1-\bar \phi)\bar \phi \phi
- r\phi(1 -\bar \phi) -\lambda(1 - \bar \phi)\bar \phi \phi^2
\nonumber \\
&+& \frac{\mathcal{A}}{8\pi D}\Big\{ \big[2\lambda\sigma(1-\bar
\phi)\bar \phi \phi - 2\lambda^2(1-\bar \phi)\bar \phi \phi^2
\big]\times
\ln \Big( \frac{2\sqrt{a_2} + a_1}{4D\mu^2} \Big) \nonumber \\
&+&  \frac{1}{16}(\Delta+ 2a_1^2) -\frac{a_1}{4}\sqrt{a_2} \Big\}
+ O(2\,\,loops).
\end{eqnarray}
The parameters appearing in this expression (and within $a_1,a_2$
and $\Delta$) are understood here to be the finite scale-dependent
ones, that is, $\sigma(\mu), r(\mu)$ and $\lambda(\mu)$. Their
explicit scale-dependence is determined below. As we mentioned
above, we have set the finite \emph{renormalized} values of $\xi$,
$b$ and $a$ identically to zero in this final expression for the
potential.

\subsection{\label{sec:GribovRGE} The one-loop RGEs}

Since $\mathcal{V}(\phi,\bar \phi)$ does not depend on the
arbitrary scale $\mu$, inserting Eq. (\ref{VGribov}) into Eq.
(\ref{RGEa}) will immediately yield the one-loop renormalization
group equations in $d=2$:
\begin{eqnarray}\label{one}
\mu \frac{d \sigma}{d \mu } &=& \frac{\mathcal{A}}{2\pi D}\sigma
\lambda
\\ \label{two}
\mu\frac{d r}{d \mu } &=& 0 + O(\mathcal{A}^2)
\\ \label{three}
 \mu\frac{d \lambda}{d \mu } &=&
\frac{\mathcal{A}}{2\pi D}\lambda^2.
\end{eqnarray}
As we saw above in Sec. (\ref{sec:pairRGE}), the one-loop RGE's
follow simply from differentiating the explicit $\mu$-dependence
in the logarithm term. The above three RGEs result from the
vanishing of the individual coefficients of the independent field
polynomials $(1-\bar \phi)\bar \phi \phi$, $(1 - \bar \phi)\phi $
and $(1 - \bar \phi)\bar \phi \phi^2$, respectively. These
equations can be solved immediately and give
\begin{eqnarray}\label{soln1}
\sigma(\mu) &=& \sigma_0
\big(\frac{\mu}{\mu_0}\big)^{\frac{\mathcal{A}\lambda_0}{2\pi D}}
\exp\big(O(\mathcal{A}^2)\big)\\
r(\mu) &=& r_0 + O(\mathcal{A}^2)\\
\label{soln3} \lambda(\mu) &=& \lambda_0 \Big(1 -
\frac{\mathcal{A}\lambda_0}{2\pi D} \ln\frac{\mu}{\mu_0} +
O(\mathcal{A}^2) \Big)^{-1}.
\end{eqnarray}
If we retain the dependence of Eq. (\ref{VGribov}) on the other
three parameters $\xi$, $b$ and $a$ and work out the consequences
implied by Eq. (\ref{RGEa}), we would find that they satisfy RGEs
similar to that for $r$ in Eq. (\ref{two}). We conclude that
neither $\xi$, $b$ nor $a$  run with scale (at one-loop order).
So, we are justified in setting $\xi(\mu) = b(\mu) = a(\mu) = 0$
\emph{after} the renormalization procedure is properly carried
out.

Although the RGE's in Eq. (\ref{one},\ref{three}) have been
calculated directly in $d=2$ dimensions, as in Sec.
(\ref{sec:pairRGE}) simple dimensional analysis can be exploited
in order to infer the \emph{approximate} form that these equations
will take in other dimensions. Dimensional analysis of the action
Eq. (\ref{sGribov}) yields (and after absorbing the diffusion
constant $D$ by a re-scaling of the time)
\begin{eqnarray}\label{dims2}
[t] &=& p^{-2} \qquad [\bar \psi] = p^0 \qquad [\psi] = p^d \qquad
[\sigma/D] = p^2, \nonumber \\
& & [\mu/D] = [r/D] = [\xi/D] = p^2 \qquad
[\lambda/D] = p^{2-d}.
\end{eqnarray}
From these relations, we are led to introduce the two
dimensionless parameters ($\epsilon = 2-d$)
\begin{equation}\label{s&g}
s = \frac{\sigma}{D\mu^2} \qquad g =
\frac{\lambda}{D\mu^{\epsilon}}.
\end{equation}
We now write Eq. (\ref{one}) and Eq. (\ref{three}) so they are
dimensionally correct, albeit only approximate in any dimension
$d$ (we herewith set $\mathcal{A} = 1$):
\begin{eqnarray}\label{oneprime}
\mu \frac{d \sigma}{d \mu } &=& \frac{K_d}{D\mu^{\epsilon}}\sigma
\lambda
\\ \label{threeprime}
 \mu \frac{d \lambda}{d \mu } &=&
\frac{K_d}{D\mu^{\epsilon}}\lambda^2,
\end{eqnarray}
where $K_d$ is defined in Eq. (\ref{solidangle}). Inserting $s$
and $g$ from Eq. (\ref{s&g}) into Eq. (\ref{oneprime}) and Eq.
(\ref{threeprime}) yields the following (approximate if $d\neq2$)
one-loop beta and flow functions
\begin{eqnarray}
\beta(g) &=& \mu\frac{d g}{d \mu } = -\epsilon g  +  K_d\, g^2 \\
\gamma(s) &=& \mu\frac{d \ln(s)}{d \mu } = -2 +  K_d\,g.
\end{eqnarray}
%

\subsection{\label{sec:Gribpotplot}Effective equation of motion}

Just as for the case of pair annihilation treated above, we can
recover the effective action from knowledge of the effective
potential and obtain the noise-corrected equations of motion for
the Gribov process. Inserting Eq. (\ref{VGribov}) into Eq.
(\ref{effactcons}), we again find that $\bar \phi = 1$ solves Eq.
(\ref{eom1}) identically while the equation of motion for $\phi$
Eq. (\ref{eom2})--the noise-averaged particle density-- works out
to be
\begin{eqnarray}\label{Gribeom}
\big(\partial_t - D\nabla^2\big)\phi &=& -\frac{\partial
\mathcal{V}}{\partial \bar{\phi}}\Big|_{\bar \phi = 1}\\ \nonumber
&=& \sigma \phi -r\phi -\lambda\phi^2 \\ \nonumber
&+& \frac{\cal{A}}{4\pi
D}(\lambda\sigma \phi - \lambda^2\phi^2) \ln\Big(\frac{r - \sigma
+2\lambda \phi}{D\mu^2}\Big) + O({\cal A}^2)\\ \nonumber &=&
F_{\phi}.
\end{eqnarray}
In Fig. \ref{fig2} we show an example of the force
experienced by the field $\phi$ for a case with $r<\sigma$ and
equilibrium density $\phi=(\sigma-r)/\lambda$. Setting
$\mathcal{A} = 0$ on the right hand side of Eq. (\ref{Gribeom})
yields the so-called rate equation which one might write down as a
first approximation to the equation of motion for the density. If
we let $n(t)$ stand for the spatial average of $\phi$, then we see
that the only stationary solutions of Eq. (\ref{Gribeom}) for zero
noise are $n = 0$, corresponding to the inactive state with no
particles remaining, and $n$ = $\frac{\sigma- r}{\lambda}$, the
active state, provided of course that $\sigma > r$. Notice that
for $\phi<(\sigma-r)/2\lambda$ the effective renormalized force is
ill-defined (it becomes complex, due to the logarithm term in the
renormalized expression). But suppose that we arrange for the
system to have initially equal rates of fission and decay: $\sigma
= r$. Then in the absence of noise, the only stationary solution
of Eq. (\ref{Gribeom}) at tree-level is the inactive state $n=0$
(see Fig 2-right, tree-level force). However, the situation
changes once the one-loop correction is included. Even when
$\sigma = r$, there can be noise-induced active states,
corresponding to finite positive values of $n$, which are given
implicitly by the transcendental equation
\begin{equation}\label{transcend}
n = \frac{(\sigma -\lambda n)}{4\pi D}\,\ln\Big(\frac{2\lambda
n}{D\mu^2}\Big).
\end{equation}
See the transitions to case 1 $(\sigma/D = 13.8)$ and to case 2
$(\sigma/D = 15.5)$. Since the physical density must be
nonnegative, for Eq. (\ref{transcend}) to have positive solutions
$n>0$ requires that either (i) $\sigma
> \lambda n > D\mu^2/2$ or (ii) $\sigma < \lambda n < D\mu^2/2$.
In the first case, we are in effect canceling a term of order
unity against a term of order $\sigma$, and there is no reason why
$\sigma$ need be small: we are not employing $\sigma$ as a
perturbation expansion parameter. Thus the finite density
solutions satisfying criteria (i) can be valid in the context of
perturbation theory. The solutions satisfying criteria (ii)
however fall prey to the same criticism as the solution given in
Eq. (\ref{newsolnpair}). In latter case, we are in effect
cancelling a term of order $\lambda$ against a term of order
$\lambda^2$.
\begin{figure}[h]
\begin{center}
\includegraphics[width=0.8  \textwidth]{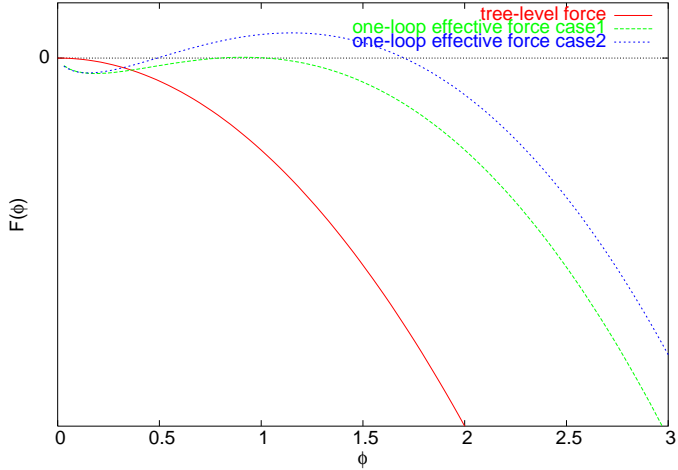}
\caption{\label{fig2} One-loop
effective and tree-level force $F(\phi)$. Depending on the values
of renormalized model parameters $\sigma(\mu), \lambda(\mu)$, the
noise can change the number of zeroes in the force from the one at
the origin (tree-level force) to two (case 1) to three (case 2).
Zeroes at finite $\phi$ signal the presence of active states.}
\end{center}
\end{figure}
%

\section{\label{sec:BARW}Branching and annihilating random walk
$ A+ A \rightarrow \emptyset$ and $A \rightarrow (m+1)A$}

In our third and final application of these calculational methods, we calculate
the renormalized one-loop effective potential for field theories
of branching and annihilating random walks (BARW). BARW describe
the dynamics of a single particle undergoing three basic
processes: diffusion, an annihilation reaction $ A+ A \rightarrow
\emptyset$ and a branching process $A \rightarrow (m+1)A$, where
$m$ is a positive integer. A field theory for these processes was
constructed and its asymptotic properties analyzed in \cite{CT},
where further references on BARW can be found. Below we deduce the
corresponding RGE's in $d=2$ dimensions for any even $m$ and then
specialize to the case $m=2$ and solve them. The associated
one-loop potential and effective force are plotted which show
qualitatively the role of small amplitude noise on the system.

\subsection{\label{sec:BARWpot} One loop effective potential}

The field theoretic action for even-$m$ BARW we will use in the
following calculations is adapted from \cite{CT} and reads as follows:
\begin{eqnarray}\label{sBARW}
S[\psi,\bar \psi] &=& \int d^dx \, dt \left[ \bar\psi
\Big(\partial_t + \xi - D\nabla^2 \Big)\psi - \lambda\big(1-
{\bar\psi}^2 \big)\psi^2 \right. \\ \nonumber
&+& \left. \sum_{l=1}^{m/2}\sigma_{2l}(1-{\bar
\psi}^{2l})\bar \psi \psi + \sum_{l=0}^{m/2}\xi_{2l}{\bar
\psi}^{2l}\right].
\end{eqnarray}
As before, $\lambda$ represents the rate of pair annihilation and
the $\sigma_{2l}$ stand for the branching rates into $2l +1$
particles, for $2l = 2,4,6,...,m$.  As pointed out in \cite{CT},
the original branching process for a given fixed integer $m$
generates all the lower branching reactions with $m-2,m-4, ...$
offspring particles by virtue of the fluctuations. All these
additional lower order branching reactions must therefore be
included in the bare action Eq. (\ref{sBARW}) in order to be able
to carry out a consistent renormalization procedure. In addition,
the terms proportional to $\xi$ and to the $\xi_{2l}$ that we have
introduced into (\ref{sBARW}) are needed to cancel off specific
quadratic and logarithmic divergences that arise in the one-loop
potential. Once we renormalize this theory, we are free to set
their finite values to zero, and we will do so below. Going
through the by now familiar steps in Eq. (\ref{smatrix}) and in
Eq. (\ref{Fourier}), we find the action Eq. (\ref{sBARW}) yields
the matrix $M(k,\omega; \phi, \bar \phi)$ of the Fourier transform
of the second order field variations, whose individual elements are given by
\begin{eqnarray}\label{matrixBARW}
M(k,\omega; \phi, \bar \phi)_{11} &=& 2\lambda\phi^2-\sum_{l=1}^{m/2}\sigma_{2l}(2l+1)(2l)\bar
\phi^{2l-1}\phi \\ \nonumber
&+& \sum_{l=0}^{m/2}\xi_{2l}2l(2l-1)\bar
\phi^{2l-2}\\
M(k,\omega; \phi, \bar \phi)_{12} &=& -i\omega + \xi + Dk^2 + 4\lambda \bar \phi \phi
+ \sum_{l=1}^{m/2}\sigma_{2l}(1-[2l+1]\bar\phi^{2l})\nonumber \\
&& \\
M(k,\omega; \phi, \bar \phi)_{21} &=& M^*(k,\omega; \phi, \bar \phi)_{12} \\
M(k,\omega; \phi, \bar \phi)_{22} &=& -2\lambda(1 - {\bar \phi}^2) .
\end{eqnarray}
Inserting this into the expression for the one-loop effective
potential Eq. (\ref{effpot}) yields
\begin{eqnarray}
\mathcal{V}(\phi,\bar \phi) &=& -\lambda\phi^2(1 - {\bar \phi}^2)
+ \sum_{l=1}^{m/2}\sigma_{2l}(1-{\bar \phi}^{2l})\bar \phi \phi +
\xi \bar \phi \phi + \sum_{l=0}^{m/2}\xi_{2l}{\bar
\phi}^{2l} \nonumber \\
&+& \frac{\mathcal{A}}{2}\int
\frac{d^dk}{(2\pi)^d}\frac{d\omega}{2\pi}\ln\Big( \frac{\det
M(k,\omega; \phi, \bar \phi)}{\det M(k,\omega;0,1)} \Big),
\end{eqnarray}
where
\begin{eqnarray}
-\det M(k,\omega; \phi, \bar \phi) = \omega^2 + \left[Dk^2 + \xi
+ 4\lambda\phi \bar \phi +
\sum_{l=1}^{m/2}\sigma_{2l}(1-[2l+1]\bar\phi^{2l})\right]^2
\nonumber
\\
+2\lambda(1-\bar\phi^2)\left[2\lambda\phi^2 -
\sum_{l=1}^{m/2}\sigma_{2l}(2l+1)(2l)\bar \phi^{2l-1}\phi +
\sum_{l=0}^{m/2}\xi_{2l}2l(2l-1)\bar \phi^{2l-2}\right],\nonumber
\end{eqnarray}
and
\begin{equation}
-\det M(k,\omega; 0,1) = \omega^2 + \big[Dk^2 + \xi -
\sum_{l=1}^{m/2}(2l)\sigma_{2l}\big]^2. \nonumber
\end{equation}}
The frequency integration is again immediate, and the remaining
(UV-regulated) integral over wavenumber $k$ in $d=2$ has
a mathematical structure identical to that in Eq. (\ref{Icutoff}),
although we need to simply replace the $(u + \xi)^2$  term by $(u
+ F)^2$ under the square-root in the second integrand, where $F =
\xi - \sum_{l=1}^{m/2}2l\sigma_{2l}$. The field dependent
coefficients appearing in $I_{\Lambda}$ in Eq. (\ref{Icutoff})
corresponding to even $m$ BARW are thus given by
\begin{eqnarray}
\label{BARWa1} a_1 &=& 2\big[\xi + 4\lambda \phi \bar \phi +
\sum_{l=1}^{m/2}\sigma_{2l}(1-[2l+1]\bar\phi^{2l})\big]
\\ \label{BARWa2}
a_2 &=& \frac{a_1^2}{4} +
2\lambda(1-\bar\phi^2)\Big[2\lambda\phi^2 -
\sum_{l=1}^{m/2}\sigma_{2l}(2l+1)(2l)\bar \phi^{2l-1}\phi \nonumber \\
&+&
\sum_{l=0}^{m/2}\xi_{2l}2l(2l - 1)\bar\phi^{2l-2} \Big] \\
\label{BARWDelta} \Delta &=& 4a_2 - a_1^2 =
8\lambda(1-\bar\phi^2)\Big[2\lambda\phi^2 -
\sum_{l=1}^{m/2}\sigma_{2l}(2l+1)(2l)\bar \phi^{2l-1}\phi \nonumber\\
&+&
\sum_{l=0}^{m/2}\xi_{2l}2l(2l - 1)\bar\phi^{2l-2}\Big].
\end{eqnarray}
Examining the ultraviolet divergences in this regulated wavenumber
integral for arbitrary even $m$, reveals the same general pattern
of quadratic and logarithmic divergences in addition to finite
terms as spelled out explicitly in Eq. (\ref{Iregulated2}). From Eq.
(\ref{BARWa1}-\ref{BARWDelta}), we immediately see that the
polynomial field structure of these two divergences contains
operators of the same order as those appearing in the tree-level
potential $V_0$ for arbitrary even $m$. The coefficients of these
quadratic and logarithmic divergences can be written as
\begin{eqnarray}\label{quaddiverge}
\big(\frac{a_1}{2} - F \big) &=& 4\lambda\bar \phi \phi +
 \sum_{l=1}^{m/2}\sigma_{2l}[2l+1](1 - \bar\phi^{2l}) \\
\frac{\Delta}{8} &=& \lambda(1-\bar\phi^2)\Big[2\lambda\phi^2 -
 \sum_{l=1}^{m/2}\sigma_{2l}(2l+1)(2l)\bar
\phi^{2l-1}\phi \nonumber \\
&+& \sum_{l=0}^{m/2}\xi_{2l}2l(2l - 1)\bar\phi^{2l-2}\Big]
\nonumber \\
&=& 2\lambda^2(1-\bar\phi^2)\phi^2 \nonumber \\
&+& \lambda
\sum_{l=1}^{m/2}[-(2l)(2l+1)\sigma_{2l} +
(2l+2)(2l+3)\sigma_{2l+2}](1 - {\bar \phi}^{2l})\bar \phi
\phi\nonumber \\ &+& 2\lambda\xi_2 + \lambda \sum_{l=1}^{m/2}
[(2l+2)(2l+1)\xi_{2l+2} - 2l(2l-1)\xi_{2l}]\bar\phi^{2l},
\end{eqnarray}
respectively, where we define $\sigma_{m+2} = 0$ and $\xi_{m+2} =
0$. For all even $m$ we conclude that this field theory is (at
least) one-loop renormalizable.  Using Eq. (\ref{useful}) once
again to separate the divergences from the finite parts in the
logarithm allows us to write the regulated one-loop potential as
follows,
\begin{eqnarray}
\mathcal{V}(\phi,\bar \phi) &=&  - \lambda(1- {\bar \phi}^2)\phi^2
+ \sum_{l=1}^{m/2}\sigma_{2l}(1-{\bar \phi}^{2l})\bar \phi \phi +
\xi\phi
\bar \phi + \sum_{l=0}^{m/2}\xi_{2l}\bar\phi^{2l}  \nonumber \\
&+& \frac{\mathcal{A}}{8\pi D}\left\{ \big(\frac{a_1}{2} - F \big)
(D\Lambda^2) + \frac{\Delta}{8}\ln \frac{\Lambda^2}{\mu^2} -
\frac{\Delta}{8}\ln\Big(\frac{2\sqrt{a_2} + a_1}{4D\mu^2} \Big)\right.
\nonumber \\
&+&
\left. \frac{1}{16}(\Delta + 2a_1^2) -\frac{a_1}{4}\sqrt{a_2} \,\right\}
+ O(\Lambda^{-2}).
\end{eqnarray}
We renormalize this potential by absorbing the divergences into
the bare parameters and introduce a set of counterterms, where
$l=(1,2,...,m/2)$
\begin{eqnarray}
\label{BARWcounter1} \lambda &=& \lambda(\mu) +
\mathcal{A}\kappa_{\lambda}
+ O(\mathcal{A}^2) \\
\label{BARWcounter2} \sigma_{2l} &=& \sigma_{2l}(\mu) +
\mathcal{A}\kappa_{\sigma_{2l}}
+ O(\mathcal{A}^2)\\
\label{BARWcounter3}
\xi &=& \xi(\mu) + \mathcal{A}\kappa_{\xi} + O(\mathcal{A}^2)\\
\label{BARWcounter4} \xi_{2l} &=& \xi_{2l}(\mu) +
\mathcal{A}\kappa_{\xi_{2l}} + O(\mathcal{A}^2)\\
\label{BARWcounter5} \xi_{0} &=& \xi_{0}(\mu) +
\mathcal{A}\kappa_{\xi_{0}} + O(\mathcal{A}^2).
\end{eqnarray}
The following choice of counterterms
\begin{eqnarray}
\kappa_{\lambda} &=& \frac{\lambda^2(\mu)}{4\pi
D}\ln \frac{\Lambda^2}{\mu^2} \\
\kappa_{\sigma_{2l}} &=&
\frac{\lambda(\mu)[2l(2l+1)\sigma_{2l}(\mu)-
(2l+2)(2l+3)\sigma_{2l+2}(\mu)]}{8\pi
D}\ln \frac{\Lambda^2}{\mu^2} \\
\kappa_{\xi} &=& -\frac{\lambda(\mu)}{2\pi}\Lambda^2 \\
\kappa_{\xi_0} &=&
-\sum_{l=1}^{m/2}\sigma_{2l}(2l+1)\frac{\Lambda^2}{8\pi}
-\frac{\lambda(\mu)\xi_2(\mu)}{4\pi
D}\ln \frac{\Lambda^2}{\mu^2} \\
\kappa_{\xi_{2l}} &=& \frac{(2l+1)\sigma_{2l}(\mu)}{8\pi}\Lambda^2\nonumber \\
&-& \frac{\lambda(\mu)}{8\pi D}[(2l+2)(2l+1)\xi_{2l+2}(\mu) -
2l(2l-1)\xi_{2l}(\mu)]\ln \frac{\Lambda^2}{\mu^2},
\end{eqnarray}
(recall $\sigma_{m+2} = \xi_{m+2} = 0$) where $l=1,2,...,m/2$,
yields the renormalized one-loop effective potential for all even
$m$ BARW:
\begin{eqnarray}\label{VBARW}
\mathcal{V}(\phi,\bar \phi) &=&  - \lambda(\mu)(1- {\bar
\phi}^2)\phi^2 + \sum_{l=1}^{m/2}\sigma_{2l}(\mu)(1-{\bar
\phi}^{2l})\bar \phi \phi + \xi\phi
\bar \phi \nonumber \\
&+& \sum_{l=0}^{m/2}\xi_{2l}(\mu)\bar\phi^{2l} \nonumber \\
&+& \frac{\mathcal{A}}{8\pi D}\left\{ -
\frac{\Delta}{8}\ln\Big(\frac{2\sqrt{a_2} + a_1}{4D\mu^2} \Big) +
\frac{1}{16}(\Delta + 2a_1^2) -\frac{a_1}{4}\sqrt{a_2} \,\right\}
\nonumber \\
&+& O(2\,\,loops).
\end{eqnarray}
Here, $a_1,a_2$ and $\Delta$ are given in Eq.
(\ref{BARWa1},\ref{BARWa2}) and Eq. (\ref{BARWDelta}),
respectively and the parameters $\lambda(\mu)$, $\sigma_{2l}(\mu)$
and $\xi_{2l}(\mu)$ appearing in this final expression are scale
dependent. We determine their scale dependence below.

-------------------------------------------------
\subsection{\label{sec:BARWRGE}The one-loop RGEs}

The renormalized potential Eq. (\ref{VBARW}) does not depend on
the arbitrary scale $\mu$.  As a result, inserting Eq.
(\ref{VBARW}) into Eq. (\ref{RGEa}) will immediately yield the
one-loop renormalization group equations in $d=2$ dimensions:
\begin{eqnarray}\label{BARWRGE1}
\mu\frac{d \lambda}{d \mu } &=& \frac{1}{2\pi D}\lambda^2\\
\label{BARWRGE2} \mu \frac{d \sigma_{2l}}{d \mu } &=&
\frac{\lambda}{4\pi D}
[2l(2l+1)\sigma_{2l}-(2l+2)(2l+3)\sigma_{2l+2}]\\
\label{BARWRGE3}
\mu\frac{d \xi}{d \mu } &=& 0 + O(\mathcal{A}^2)\\
\label{BARWRGE4}
\mu\frac{d \xi_0}{d \mu } &=& -\frac{\lambda \xi_2}{2\pi D}\\
\label{BARWRGE5} \mu\frac{d \xi_{2l}}{d \mu } &=&
\frac{\lambda}{4\pi D} [2l(2l-1)\xi_{2l}-(2l+2)(2l+1)\xi_{2l+2}].
\end{eqnarray}
As we saw above in Sec. (\ref{sec:pairRGE}) and in Sec.
(\ref{sec:GribovRGE}), at one-loop order, the RGE's follow
immediately from differentiating the explicit $\mu$-dependence in
the logarithm term in Eq. (\ref{VBARW}). The above set of RGEs
results from picking off the coefficients of the field polynomials
$(1-{\bar \phi}^2){\phi}^2$, $(1 - {\bar \phi}^{2l})\bar \phi
\phi$, for $l=1,2,...,m/2$, $\bar \phi \phi$ and of ${\bar
\phi}^{2l}$ , for $l=0,1,2,...,m/2$, respectively. Note that the
RGE Eq. (\ref{BARWRGE2}) at the highest value $l=m/2$ is given by
\begin{equation}\label{mRGE}
\mu \frac{d \sigma_{m}}{d \mu} = \frac{m(m+1)}{4\pi D}\lambda
\sigma_{m}.
\end{equation}
Up to this stage, we have carried through the complete calculation
of the effective potential valid for arbitrary even $m$. However,
it is known from Ref. \cite{CT} that the most relevant of the
branching reactions is actually the one with smallest $m$, namely
$m=2$. The branching process with $m=2$ therefore will describe
the entire universality class of BARW with even offspring. For
this reason, we now specialize the remainder of this section to
this case.

The explicit solutions of the $m=2$ RGE's Eq.
(\ref{BARWRGE1}-\ref{BARWRGE5}) are as follows:
\begin{eqnarray}
\lambda(\mu) &=& \lambda_0 \Big(1 -
\frac{\mathcal{A}\lambda_0}{2\pi D}\ln \frac{\mu}{\mu_0}
+ O(\mathcal{A}^2)\Big)^{-1}\\
\sigma_2(\mu) &=& \sigma_2(0)
\Big(\frac{\mu}{\mu_0}\Big)^{\frac{3\mathcal{A}\lambda_0}{2\pi
D}}\exp \big(O(\mathcal{A}^2)\big)\\
\xi(\mu) &=& \xi(0) + O(\mathcal{A}^2) \\
\xi_0(\mu) &=& \xi_0(0) - \frac{\mathcal{A}\lambda_0\xi_2(0)}{2\pi
D}\ln \frac{\mu}{\mu_0} + O(\mathcal{A}^2)\\
\xi_2(\mu) &=&
\xi_2(0)\Big(\frac{\mu}{\mu_0}\Big)^{\frac{\mathcal{A}\lambda_0}{2\pi
D}}\exp \big(O(\mathcal{A}^2)\big).
\end{eqnarray}
According to these solutions, we can now set the renormalized
values of $\xi(\mu),\xi_0(\mu)$ and $\xi_2(\mu)$ all identically
equal to zero at all scales $\mu$, by simply choosing their values
at the arbitrary initial scale $\mu_0$ to be zero: $\xi(0) =
\xi_0(0)= \xi_2(0) = 0$. Our final renormalized potential depends
only on the rate of pair annihilation and the branching rate. The
potential and effective force will be calculated with this choice
(see Sec(\ref{sec:BARWplot}) below).

The RGE's Eq. (\ref{BARWRGE1}-\ref{BARWRGE5}) have been calculated
directly in $d=2$, which coincides the upper critical dimension
for BARW. As before, we can deduce crude and approximate forms
that these RGE's must take in any dimension by combining simple
dimensional analysis with these $d=2$ RGE's. We take Eq.
(\ref{BARWRGE1}) and Eq. (\ref{mRGE}) for $m=2$ and write them as
follows:
\begin{eqnarray}\label{dimextendeda}
\mu\frac{d \lambda}{d \mu } &=& \frac{1}{2\pi D \mu^{\epsilon}}\lambda^2\\
\label{dimextendedb} \mu \frac{d \sigma_{2}}{d \mu} &=&
\frac{3}{2\pi D \mu^{\epsilon}}\lambda \sigma_{2}.
\end{eqnarray}
Then introduce the dimensionless parameters
\begin{eqnarray}
g &=& \frac{K_d \lambda}{D \mu^{\epsilon}},\\
s &=& \frac{\sigma_2}{D \mu^2},
\end{eqnarray}
into Eq. (\ref{dimextendeda},\ref{dimextendedb}). This yields the
approximate one-loop beta and zeta functions for $m=2$ BARW in
dimension $d$:
\begin{eqnarray}
\beta(g) &=& \mu\frac{d g}{d \mu } = -\epsilon g + \frac{1}{2\pi K_d}\,g^2,\\
\zeta(s) &=& \mu\frac{d \ln(s)}{d \mu } = -2 + \frac{3}{2\pi
K_d}\,g.
\end{eqnarray}
For $d=2$, these reduce to
\begin{eqnarray}
\beta(g) &=& \mu\frac{d g}{d \mu } = g^2,\\
\zeta(s) &=& \mu\frac{d \ln(s)}{d \mu } = -2 + 3g,
\end{eqnarray}
and reproduce the $d=2$ one-loop beta and zeta functions obtained
in \cite{CT} by other means.

\subsection{\label{sec:BARWplot}Effective equation of motion}

\begin{figure}[h]
\begin{center}
\includegraphics[width=0.8\textwidth]{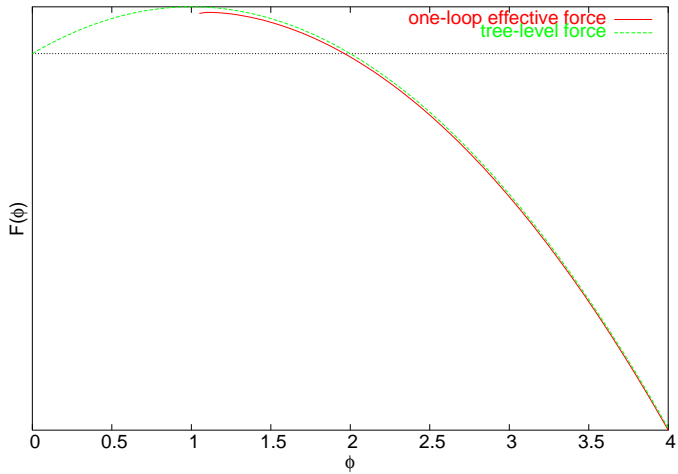}
\caption{\label{fig3} One-loop
effective and tree-level force $F(\phi)$. The zero at the origin
corresponds to the inactive state, while the zero at finite $\phi$
corresponds to an active state.}
\end{center}
\end{figure}

We can calculate the effective force and equation of motion for
the physical density $\phi$ starting from Eq. (\ref{effactcons})
(note there is no wavefunction renormalization for BARW
\cite{CT}). Once again, we find that $\bar \phi = 1$ solves
identically its equation of motion Eq. (\ref{eom1}), while
evolution of $\phi$ is determined by by Eq. (\ref{eom2}), which
when written out yields
\begin{eqnarray}\label{BARWeom}
\big(\partial_t - D\nabla^2\big)\phi &=&  -\frac{\partial
\mathcal{V}}{\partial \bar{\phi}}\Big|_{\bar \phi = 1}\nonumber \\
\nonumber &=& 2\sigma_2 \phi -2\lambda\phi^2 \nonumber \\
&+& \frac{\cal{A}}{2\pi
D}\lambda(3\sigma_2 \phi - \lambda \phi^2) \ln\Big(\frac{4\lambda
\phi - 2\sigma_2}{D\mu^2}\Big) + O({\cal A}^2) \nonumber \\
&=&
F_{\phi}.
\end{eqnarray}
In Fig. \ref{fig3} we show an example of the force
$F_{\phi}$ experienced by the field $\phi$ with zeroes at
$\phi=\sigma_2/\lambda$. The one-loop force is ill defined when
the argument of the logarithm goes negative. Here, if we set
directly $\sigma_2 = 0$ in the force in Eq. (\ref{BARWeom}), the
one-loop correction does not induce any active states that can be
considered valid within perturbation theory, for the same reasons
as spelled out in the case of pair-annihilation.

----------------------------------------------------------------
\section{\label{sec:disc}Discussion}

In this paper we have taken the well established effective action
and effective potential formalism from the quantum field theory
domain and have adapted and applied it to classical field theories
of the Doi and Peliti type of reacting and diffusing particles.
These are field-theoretic representations of the corresponding
diffusion-limited reaction problems derived from classical master
equations. Apart from taking the continuum limit, the resultant
actions require no further approximation and the noise is included
exactly. The main result of our application is the formula for the
one-loop effective potential which holds for any multiple
$N$-species reaction-diffusion system subject to internal particle
density fluctuations. The main goal of this paper is to illustrate
the relative ease with which this effective potential can be
analytically calculated in closed form and consistently
renormalized for a wide variety of interesting single ($N=1$)
species models, of increasing complexity, from the simplest pair
annihilation case to the more intricate case of branching and
annihilating random walks with arbitrary even number $m$ of offspring.
We believe the most interesting specific application of the
effective potential formalism in these problems is the ease with
which one can deduce the associated one-loop renormalization group
equations (RGE) and the effective equations of motion from it.

We have calculated the effective potential in fixed space
dimension $d=2$, corresponding to the upper critical dimension for
all the models surveyed here. Once the frequency integration is
carried out in Eq. (\ref{effpot}), we have seen that the remaining
integral over the wavevector always reduces to the same small set
of simple elementary functions (polynomials and logarithms),
irrespective of the microscopic details of the specific model. This feature is
what allows one to obtain potentials and the RGE's for a wide
variety of different models with little additional effort. In so
far as this is a consequence of the one-loop approximation, this
does represent a major technical advantage over the standard
Feynman diagram approach, since in the latter approach, each
individual model requires the handling of a distinct set of
diagrams with its attendant combinatorics to work out. Of course,
when one loop order is not sufficient, to go to higher loops, one
must revert back to diagrams to compute the higher order
corrections to the potential, and this can be done \cite{Jackiw}.
While individual Feynman diagrams can be analytically continued to
arbitrary dimension, this appears to be more difficult for the
potential, representing, as it does, an infinite summation of
diagrams \cite{CW}. Nevertheless, the analytic continuation of the
one-loop potential to arbitrary dimensions may be possible, and
this point clearly deserves further work. This would open up the
very interesting possibility to derive the associated RGE's for a
wide variety of reaction-diffusion models in a straightforward
manner for any dimension $d$ in a universal way, at one-loop
level, without the need to handle individual Feynman diagrams. It
would be interesting to test out these analytic methods on
two-species ($N=2$) reaction-diffusion systems such as the $A+B
\rightarrow 0$ reaction \cite{LeeCardy}. The expression for an $N=2$
species potential requires working out the determinant of
$4 \times 4$ arrays, followed by integrations over frequency and
wavenumber.
Our conviction is that further developments of these potential
methods can promote significant advances in nonequilibrium
statistical mecanics and in physical chemistry. The simple
examples treated here serve as a testing ground for this
technique, and we believe we have amply demonstrated the
usefulness of the approach and the relative ease in carrying out
the specific steps leading to the renormalization group equations.

\appendix
\section{Effective potential}
In this appendix we derive the one-loop effective potential Eq. (\ref{effpot}) for
Doi-Peliti type stochastic field theories. Following standard
procedures \cite{Doi,Peliti}, one starts from the microscopic
master equation corresponding to the specific chemical reactions
to be studied and then passes to a field-theoretic solution of
this master equation. In the final step, one obtains a path
integral representation of the time-evolution operator (or
``hamiltonian") of the general form
\begin{equation}\label{pathintegral}
Z \sim \int\Pi^N_{j} (\mathcal{D}\psi_j \mathcal{D}\bar \psi_j)
\exp\Big(-S[\psi_i, \bar \psi_i]\Big),
\end{equation}
for ${N}$-species of particles $\psi_i$ and ${N}$ conjugate or
``response" fields $\bar \psi_i$ for $i = (1,2,\ldots,N)$, and
where $S$ denotes the unrenormalized (bare) action.  Introduce the
loop counting parameter $\mathcal{A}= 1$ in the exponential in Eq.
(\ref{pathintegral}) as follows
\begin{equation}\label{loop}
S[\psi_i, \bar \psi_i] \rightarrow S[\psi_i, \bar
\psi_i]/\mathcal{A}.
\end{equation}
Fluctuations are taken into account via the loop expansion in
powers of $\mathcal{A}$. Since the loop expansion corresponds to
an expansion in a parameter that multiplies the entire action, it
is unaffected by shifts of fields.

We next adapt the basic steps that take us from an action $S$ to
the effective action and effective potential \cite{
HPMVb,HPMVa,Jackiw,IIM,ZJ,Rivers,CW}. First, we introduce a set of
arbitrary source functions $J_i$ and $\bar J_i$ for both the
$\psi_i$ and $\bar \psi_i$ by adding the following term to the
action
\begin{equation}\label{sources}
S[\psi_i, \bar \psi_i] \rightarrow S[\psi_i,\bar \psi_i] + \int
d^dx \, \int dt\, \sum_{i=1}^N(J_i\psi_i + {\bar J}_i{\bar
\psi}_i),
\end{equation}
this linear shift leads to a source dependence in the path
integral Eq. (\ref{pathintegral}) which we denote by $Z[J,{\bar
J}]$. Next, introduce the functional for connected correlation
functions $W[J,{\bar J}]$ and its Legendre transform, the
effective action $\Gamma[\phi, {\bar \phi}]$
\begin{eqnarray}\label{W}
W[J,{\bar J}] &=& \mathcal{A}\ln Z[J,{\bar J}],\\\label{Gamma}
\Gamma[\phi, {\bar \phi}] &=& -W[J,{\bar J}] + \int d^dx \, \int
dt\, \sum_{i=1}^N(J_i\phi_i + {\bar J}_i{\bar \phi}_i),
\end{eqnarray}
where
\begin{equation}\label{defphi}
\phi_i[J,{\bar J}] = \frac{\delta W[J,\bar J]}{\delta J_i} ,
\qquad {\bar \phi_i}[J,{\bar J}] = \frac{\delta W[J,\bar
J]}{\delta {\bar J}_i}.
\end{equation}
The fields $\phi_i$ and $J_i$ (and $\bar \phi_i$ and $\bar J_i$)
make up the Legendre transform pairs. From Eq.
(\ref{pathintegral},\ref{sources},\ref{W}), and Eq.
(\ref{defphi}), we see that the fields $\phi_i,\bar \phi_i$ are
the fluctuation averages of the original degrees of freedom, but
evaluated in the presence of the source terms, that is,
\begin{equation}\label{averages}
\phi_i[J,{\bar J}] = \langle \psi_i(x,t) \rangle_{J,{\bar J}}
\qquad {\bar \phi}_i[J,{\bar J}] = \langle \bar \psi_i(x,t)
\rangle_{J,{\bar J}}\,,
\end{equation}
where the angular brackets denote the average over the
fluctuations. These fluctuation-averaged fields are the solutions
of the \emph{effective} equations of motion
\begin{equation}\label{effeqomJ}
\Big(\frac{\delta \Gamma[\phi, \bar \phi]}{\delta
\phi_i}\Big)\Big|_{\phi[J,\bar J],\bar \phi[J,\bar J]} = J_i,
\qquad \Big(\frac{\delta \Gamma[\phi, \bar \phi]}{\delta \bar
\phi_i}\Big)\Big|_{\phi[J,\bar J],\bar \phi[J,\bar J]} = {\bar
J}_i,
\end{equation}
in the presence of the source terms, a fact which follows
immediately from differentiating Eq. (\ref{Gamma}) as indicated.
Most importantly, once we set the external sources to zero in Eq.
(\ref{effeqomJ}), $J,\bar J = 0$, we then obtain the effective
equations of motion for the stochastic-averaged fields. The
solutions are the stationary points of the effective action:
\begin{equation}\label{effeqom0}
\Big(\frac{\delta \Gamma[\phi, \bar \phi]}{\delta
\phi_i}\Big)\Big|_{\phi[0,0],\bar \phi[0,0]} = 0, \qquad
\Big(\frac{\delta \Gamma[\phi, \bar \phi]}{\delta \bar
\phi_i}\Big)\Big|_{\phi[0,0],\bar \phi[0,0]} = 0.
\end{equation}
We make use of Eq. (\ref{effeqom0}) in the paper in specific
reaction diffusion models and explicitly obtain the
fluctuation-modified effective equations of motion satisfied by
the noise-averaged particle densities at one-loop order.

To complete the derivation of the one-loop effective potential
(and suppressing writing out the full dependence on indices, and
spatial and temporal arguments) we henceforth let $(\phi,\bar
\phi)$ denote the set of the source-free fluctuation averages of
the $2N$ scalar fields $\{ \langle \psi_i\rangle_{(0,0)},
\langle\bar \psi_i\rangle_{(0,0)} \}_{i = 1}^N$. Then the
effective action for the field theory described by Eq.
(\ref{pathintegral}) is given by \cite{ZJ}
\begin{equation}\label{effaction}
\Gamma[\phi,{\bar \phi}] = S[\phi,{\bar \phi}] +
\mathcal{A}\Gamma_1[\phi,{\bar \phi}] + O(\mathcal{A}^2),
\end{equation}
where the one-loop contribution is explicitly calculated to be
\begin{equation}\label{oneloop}
\Gamma_1[\phi,{\bar \phi}] =
\frac{1}{2}\mathrm{Tr}[\mathrm{Ln}\,S^{(2)}(\phi,{\bar \phi};x,y)
- \mathrm{Ln}\,S^{(2)}(0,1;x,y)],
\end{equation}
and $\mathrm{Tr}$ and $\mathrm{Ln}$ are functional traces and
logarithms (these operate on both discrete and continuous
indices). The array $S^{(2)}$ of second order functional
derivatives of the action with respect to the fields is expressed
in $2\times2$ block array form in Eq. (\ref{smatrix}), where $i,j
= (1,2,\ldots,N)$. The action $S[\phi,{\bar \phi}]$ is simply the
action Eq. (\ref{loop}) after replacing $\psi_i$ and $\bar \psi_i$
by $\phi_i$ and $\bar \phi_i$, respectively. The effective
potential is the effective action Eq. (\ref{effaction}) evaluated
on \emph{constant} field configurations, and after dividing out by
the overall volume of space-time $VT$:
\begin{eqnarray}\label{definition}
\mathcal{V}[\phi,{\bar \phi}] \equiv \frac{\Gamma[\phi,{\bar
\phi}]}{VT} &=& \frac{S[\phi,{\bar \phi}]}{VT} +
\mathcal{A}\frac{\Gamma_1[\phi,{\bar \phi}]}{VT} +
O(\mathcal{A}^2)\nonumber \\ &=& V_0[\phi,{\bar \phi}] +
\mathcal{A} V_1[\phi,{\bar \phi}] + O(\mathcal{A}^2).
\end{eqnarray}
For constant fields $\phi,{\bar \phi} = \textrm{const.}$, we pass
to Fourier space and define a $2N \times 2N$ matrix $M$ via the
Fourier transform (FT) of Eq. (\ref{smatrix}) as defined in Eq.
(\ref{Fourier}). Since this expression is diagonal in momentum and
frequency, we can straightforwardly carry out the sequence of
operations indicated in Eq. (\ref{oneloop}) \cite{Jackiw}:
\begin{eqnarray}\label{tracelogM}
& &\mathrm{Tr}\mathrm{Ln}[M(k,\omega; \phi,{\bar \phi})\times
(2\pi)^{d+1}\delta^d(k-p)\delta(\omega - \Omega)] \nonumber \\
&=& \mathrm{Tr}(2\pi)^{d+1}\delta^d(k-p)\delta(\omega - \Omega)\ln
M(k,\omega,\phi,\bar \phi)\nonumber \\
&=& \int \frac{d^dk}{(2\pi)^d}
\int\frac{d\omega}{2\pi}(2\pi)^{d+1}\delta^d(k-p) \delta(\omega -
\Omega)|_{k=p,\omega =\Omega}\,
\mathrm{tr}\ln M(k,\omega,\phi,\bar \phi)\nonumber \\
&=& \int d^dx \, \int dt\, \int \frac{d^dk}{(2\pi)^d}
\int\frac{d\omega}{2\pi} \ln \det M(k,\omega,\phi,\bar \phi),
\end{eqnarray}
where $\ln$ and $\mathrm{tr}$ are the ordinary logarithm and
trace, and we have used the identity $\mathrm{tr}\ln M = \ln \det
M$.
Evaluating Eq. (\ref{effaction}) on constant configurations and
dividing through by the volume of space-time and using Eq.
(\ref{smatrix},\ref{Fourier},\ref{oneloop},\ref{definition}) and
Eq. (\ref{tracelogM}), yields the compact expression for the
one-loop effective potential Eq. (\ref{effpot}).

\flushleft{\textbf{Acknowledgements}}\\
M.-P.Z. acknowledges a fellowship provided by INTA for training in
astrobiology. The research of D.H. is supported in part by funds
from CSIC and INTA.


\begin{thebibliography}{99}
\bibitem{HPMVb} D. Hochberg, C.
Molina-Par\'{\i}s, J. P\'erez-Mercader and M. Visser, Effective
action for stochastic partial differential equations, Phys. Rev. E
\textbf{60}:6343-6360 (1999).
\bibitem{Oerding} K. Oerding, F. van Wijland, J.-P. Leroy and H.J.
Hilhorst, Fluctuation-Induced First-Order Transition in a
Nonequilibrium Steady State, J. Stat. Phys. \textbf{99}:1365-1395
(2000).
\bibitem{HK} H.K. Janssen, \"{U}. Kutbay and K. Oerding, Equation of state for directed
percolation, J. Phys. A: Math. Gen. \textbf{32}:1809-1817 (1999).
\bibitem{Canet} L. Canet, B. Delamotte, O. Deloubri\`{e}re and N. Wschebor,
Nonperturbative Renormalization-Group Study of Reaction-Diffusion
Processes , Phys. Rev. Lett. \textbf{92}:195703(1)-195703(4)
(2004); L. Canet, H. Chat\'{e} and B. Delamotte, Quantitative
Phase Diagrams of Branching and Annihilating Random Walks, Phys.
Rev. Lett. \textbf{92}:255703(1)-255703(4) (2004).
\bibitem{Doi} M. Doi, Second quantization representation for
classical many-particle system, J. Phys. A: Math. Gen.
\textbf{9}:1465-1477 (1976).
\bibitem{Peliti} L. Peliti, Path integral approach to birth-death processes
on a lattice, J. Physique \textbf{46}:1469-1483 (1985).
\bibitem{HPMVa} D. Hochberg, J. P\'erez-Mercader, C.
Molina-Par\'{\i}s and M. Visser, Renormalization Group Improving
the Effective Action: A Review, Int. J. Mod. Phys. A \textbf{14}:
1485-1521 (1999).
\bibitem{FRP} Y. Fujimoto, L. O'Raifeartaigh and G. Parravicini,
Effective potential for non-convex potentials, Nucl. Phys.
B\textbf{212}:268-300 (1983).
\bibitem{GLPQ} B. Gato, J. Leon, J. P\'{e}rez-Mercader and M.
Quir\'{o}s, Renormalization group analysis for a general softly
broken supersymmetric gauge theory, Nucl. Phys.
B\textbf{253}:285-307 (1985).
\bibitem{TW} D. Toussaint and F. Wilczek, Particle-antiparticle annihilation
in diffusive motion, J. Chem. Phys. \textbf{78}:2642-2647 (1983).
\bibitem{KR} K. Kang and S. Redner, Scaling approach for the Kinetics of
Recombination Processes, Phys. Rev. Lett. \textbf{52}:955-958
(1984).
\bibitem{Peliti86} L. Peliti, Renormalisation of fluctuation effects in the
$A+A \rightarrow A$ reaction, J. Phys. A: Math. Gen.
\textbf{19}:L365-L367 (1986).
\bibitem{Ohtsuki} T. Ohtsuki, Field-theoretical aproach to scaling behavior
of diffusion-controlled recombination, Phys. Rev. A\textbf{43}:
6917-6919 (1991).
\bibitem{DrozSas} M. Droz and L. Sasv\'{a}ri, Renormalization-group approach to simple
reaction-diffusion phenomena, Phys. Rev. E\textbf{48}:R2343-R2346
(1993).
\bibitem{FLS}B. Friedman, G. Levine and B. O'Shaughnessy, Renormalization-group
study of field-theoretic $A+A \rightarrow 0$, Phys. Rev.
A\textbf{46}: R7343-R7346 (1992).
\bibitem{BPLee} B.P. Lee, Renormalization group calculation for the
reaction $kA \rightarrow 0$, J. Phys. A: Math Gen.
\textbf{27}:2633-2652 (1994).
\bibitem{GR} I.S. Gradshteyn and I.M. Ryzhik, \textit{Table of Integrals,
Series and Products }(Academic Press, New York, 1980).
\bibitem{Jackiw} R. Jackiw, Functional evaluation of the effective potential,
Phys. Rev. D\textbf{9}:1686-1701 (1974).
\bibitem{IIM} J. Iliopoulos, C. Itzykson, A. Martin, Functional methods and
perturbation theory, Rev. Mod. Phys. \textbf{47}:165-192 (1975).
\bibitem{ZJ} J. Zinn-Justin, \textit{Quantum Field Theory and Critical
Phenomena} (Oxford University Press, Oxford, 2002) 4rth edition.
\bibitem{Rivers} R.J. Rivers, \textit{Path integral methods in
quantum field theory} (Cambridge University Press, Cambridge,
1988).
\bibitem{CW} S. Coleman and E. Weinberg, Radiative Corrections as the Origin of
Spontaneous Symmetry Breaking, Phys. Rev. D\textbf{7}:1888-1910
(1973).
\bibitem{GS} P. Grassberger and K. Sundermeyer, Reggeon field theory and Markov
processes, Phys. Lett. \textbf{77}B:220-222 (1978).
\bibitem{J} H.K. Janssen, On the Nonequilibrium Phase Transition in
Reaction-Diffusion Systems with an Absorbing Stationary State, Z.
Phys. B\textbf{42}:151-154 (1981).
\bibitem{Howard} M.J. Howard and U.C. T\"{a}uber, Real versus imaginary noise
in diffusion-limited reactions, J. Phys. A: Math. Gen.
\textbf{30}:7721-7731 (1997).
\bibitem{CT} J.L. Cardy and U.C. T\"{a}uber, Field Theory of Branching
and Annihilating Random Walks, J. Stat. Phys. \textbf{90}:1-56
(1998).
\bibitem{LeeCardy} B.P. Lee and J. Cardy, Renormalization Group Study of the
$A + B \rightarrow 0$ Diffusion-Limited Reaction, J. Stat. Phys.
\textbf{80}:971-1007 (1995); \textit{ibid.} \textbf{87}:951-954
(1997).
\end{thebibliography}
\end{document}